\documentclass[apl, twocolumn,tightenlines,superscriptaddress]{revtex4-1}

\usepackage{amsthm}
\usepackage{amsmath}
\usepackage{bbm, dsfont}
\usepackage{graphicx}
\usepackage[usenames,dvipsnames]{color}
\usepackage[colorlinks=true,citecolor=magenta,urlcolor=blue]{hyperref}
\usepackage[table]{xcolor}
\usepackage{colortbl}
\usepackage{color}
\usepackage{multirow}
\usepackage{hhline}
\usepackage{tabu}
\usepackage{epstopdf}
\usepackage{booktabs}
\usepackage[normalem]{ulem}
\usepackage{siunitx}
\usepackage{enumitem}
\usepackage{braket}
\usepackage{siunitx}
\usepackage{textcomp}
\usepackage{booktabs}
\usepackage[textwidth=3cm, textsize=footnotesize]{todonotes}
\usepackage[font=small,skip=3pt]{caption}

\DeclareSIUnit\permille{\text{\textperthousand}}

\graphicspath{{./Figures/}}

\newcolumntype{C}[1]{ >{ \centering\arraybackslash}p{#1}}

\definecolor{gray4}{gray}{0.8}
\definecolor{gray2}{gray}{0.6}

\begin{document}
\title{ Performance and security of 5~GHz repetition rate polarization-based Quantum Key Distribution} 

\author{Fadri Gr\"unenfelder}\email{fadri.gruenenfelder@unige.ch}
\affiliation{Group of Applied Physics, University of Geneva, Chemin de Pinchat 22, CH-1211 Geneva 4, Switzerland}
\author{Alberto Boaron}
\affiliation{Group of Applied Physics, University of Geneva, Chemin de Pinchat 22, CH-1211 Geneva 4, Switzerland}
\author{Davide Rusca}
\affiliation{Group of Applied Physics, University of Geneva, Chemin de Pinchat 22, CH-1211 Geneva 4, Switzerland}
\author{Anthony Martin}
\affiliation{Group of Applied Physics, University of Geneva, Chemin de Pinchat 22, CH-1211 Geneva 4, Switzerland}
\affiliation{ {\normalfont Now at:} Universite Cote d’Azur, CNRS, Institut de Physique de Nice, Parc Valrose, 06108 Nice Cedex 2, France}
\author{Hugo Zbinden}
\affiliation{Group of Applied Physics, University of Geneva, Chemin de Pinchat 22, CH-1211 Geneva 4, Switzerland}

\begin{abstract}
We present and characterize a source for a \SI{5}{\giga\hertz} clocked polarization-based simplified BB84 protocol. Secret keys are distributed over \SI{151.5}{\kilo\meter} of standard telecom fiber at a rate of 54.5 kbps. Potentially, an increased clock frequency of the experiment introduces correlations between succeeding pulses.
We discuss the impact of these correlations and propose measurements to estimate the relevant parameters.
\end{abstract}

\maketitle


Quantum key distribution (QKD) offers a way to distribute a secret key over a distance between two communicating parties, Alice and Bob. Since the first experimental demonstration of the BB84 protocol in 1992 \cite {Bennett1992} huge progress has been achieved and commercial devices are now available. 
QKD systems with up to $\SI{2.5}{\giga\hertz}$ repetition rate were used to exchange secret keys both at high rates \cite{Yuan2018} and over long distances \cite{Boaron2018,Wang2012}. Furthermore, early experiments with repetition rates of up to $\SI{10}{\giga\hertz}$ using the differential phase shift protocol \cite{Takesue2007b,Choi2010} were performed.

However, none of the previously mentioned studies considered the security issues which can appear due to the increased repetition rate. Indeed, several  issues concerning correlations between succeeding pulses have been discussed in literature. E.g. Kobayashi \textit{et al.}  modelled the phase distribution of a gain-switched lasers used in QKD experiments \cite{Kobayashi2014} and showed how to characterize phase randomization. Other articles investigated correlations which appear due to imperfect decoy encoding and gave ways to prevent those \cite{Roberts2018, Yoshino2018}. 

In this work, we demonstrate secret key exchange using the simplified BB84 protocol  \cite{Tamaki2014, Pereira2019, Gruenenfelder2018} and polarization encoding running at a pulse repetition rate of \SI{5}{\giga\hertz} over distances of \SI{101}{\kilo\meter} and \SI{151.5}{\kilo\meter}. Moreover, we identify three possible security loopholes which are related to the high repetition rate and measure the corresponding parameters. The first potential loophole may arise when encoding the qubit, since the polarization states of succeeding pulses may be slightly correlated. Similarly a correlation may appear in the intensities of the decoy and signal states \cite{Yoshino2018}. The third loophole concerns the laser source. A necessary assumption in security proofs of decoy state BB84 is a random phase between the pulses  emitted by the gain-switched distributed feedback lasers \cite{Ma2005}. However, at high pulse rates the phase correlations between the pulses may appear \cite{Kobayashi2014}. We characterize the degree of phase-randomization by means of an interference experiment and discuss a model to take non-perfect phase-randomization into account when calculating the secret key rate (SKR). Typically, these three loopholes become stronger at high pulse repetition rate and it is important to characterize these effects.


\begin{figure*}
\includegraphics[ trim={0 0 0 0},clip, width = 1.9\columnwidth]{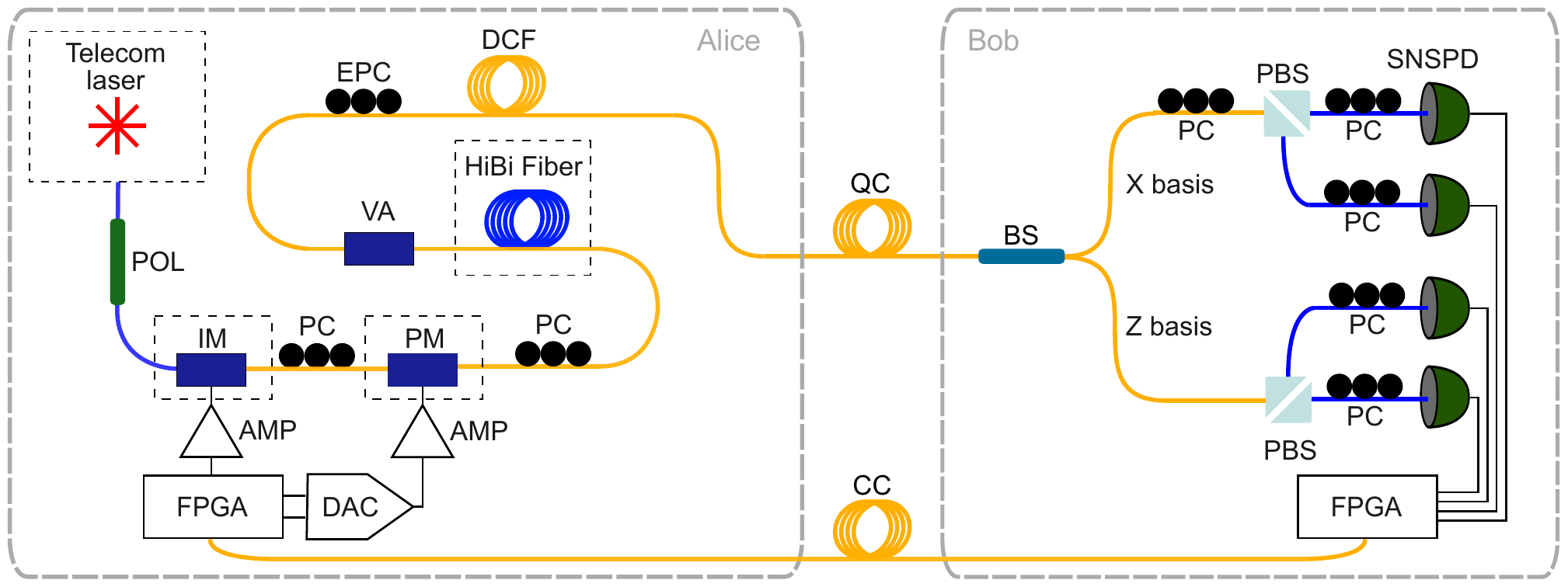}
\caption{\label{fig:setup} Schematic of the QKD setup. AMP: electrical amplifier; BS: beamsplitter; CC: classical channel; DAC: digital-to-analog converter; DCF: dispersion compensating fiber; EPC: electronic polarization controller; FPGA: field programmable gate array; HiBi Fiber: highly birefringent fiber; IM: intensity modulator; PBS: polarizing beamsplitter; PC: polarization controller; PM: phase modulator; POL: polarizer; QC: quantum channel; SNSPD: Superconducting nanowire single photon detector; VA: variable attenuator. The dashed boxes are temperature stabilized. The blue lines represent highly birefringent fibers. Electric wiring is shown as black lines.}
\end{figure*}

We employ the fiber-based setup depicted in \autoref{fig:setup} and  implement the 1-decoy protocol \cite{Ma2005,Rusca2018}. Alice uses a gain-switched distributed feedback multi-quantum well laser (Gooch and Housego AA0701) driven at a repetition rate of \SI{5}{\giga\hertz} to create pulses at a center wavelength of \SI{1550.92}{\nano\meter} with a duration of \SI{43}{\pico\second} (FWHM). After an additional polarizer (POL), the light is polarized with an extinction of more than \SI{40}{\decibel}. A Lithium Niobate electro-optical intensity modulator (IM) with a bandwidth of \SI{20}{\giga\hertz} (Oclaro SD-20) is used to switch between signal and a decoy intensites. The polarization modulation is achieved via an electro-optical phase modulator (PM) with a bandwidth of \SI{10}{\giga\hertz} (iXblue MPZ-LN-10). The laser pulses enter the PM with a linear polarization at an angle of \SI{45}{\degree} with respect to the fast axis of the birefringent crystal inside the PM. A voltage applied to the crystal changes the difference in refractive index between the fast and slow axis and therefore changes the relative phase between the two field components. 
The PM has a nominal  half-wave voltage $V_\pi$ of $\SI{7}{\volt}$, i.e. the voltage needed to delay a pulse  along a specific axis by a phase of $\pi$ (or $\lambda/2$)). However, the electric field affects also the orthogonal axis. This means in order to obtain a phase of $\pi$ between the orthogonal polarization modes a voltage higher than $\SI{7}{\volt}$ is needed. A field programmable gate array (FPGA) controls both  the IM and PM. Since the IM requires only two levels, we drive it by directly amplifying the digital signal of the FPGA. To drive the PM, a digital-to-analog converter (DAC) with a symbol rate up to 32 GBaud (SHF 611 F) is used to provide a three-level signal.

The birefringence of the PM crystal induces polarization mode dispersion which is compensated by a piece of highly birefringent fiber with an approximate length of \SI{9}{\meter}. A variable attenuator is used to reduce the mean photon numbers at the output of Alice of the signal and decoy state to $\mu_0$ and  $\mu_1$, respectively. With an electronic polarization controller Alice stabilizes the polarization arriving at Bob by using a feedback loop which maximizes the SKR. The disperion compensating fiber is used to compensate the chromatic dispersion of the quantum channel. Bob detects the states with in-house made superconducting nanowire single photon detectors (SNSPD) based on molybdenum silicide \cite{Caloz2018}.
The efficiency of these detectors varies with the incident polarization. To obtain the same efficiency for all four states, we align the polarization of the incoming pulses just in front of the detectors.

The protocol goes as follows: Alice prepares randomly one of three possible states. Either she sends one of the two orthogonal states $\ket{0}$ and $\ket{1}$ which form the Z basis or she sends $\ket{+} = (\ket{0}+\ket{1})/\sqrt{2}$. The states  $\ket{+}$ and $\ket{-} = (\ket{0}-\ket{1})/\sqrt{2}$ form the X basis. The random numbers used for the choice of the state are generated in real time using 26 AES cores seeded by a quantum random number generator \cite{Boaron2018}. Bob chooses between the Z and X basis passively with a balanced beam splitter. The secret key is extracted from outcomes where both Alice and Bob choose the Z basis and all other events can be used to estimate the information a potential eavesdropper Eve has on the key.


We perform secret key exchanges over spooled fibers of  \SI{101.0}{\kilo\meter} and \SI{151.5}{\kilo\meter} length. The chosen mean photon numbers for the decoy states are $\mu_0 = 0.3$ and $\mu_1 = 0.15$ and the probability of sending $\mu_0$ is $0.6$ for both distances. These values were found by optimizing the SKR numerically. The probability that Alice sends the Z-basis is $0.9$ and the probability of Bob choosing the Z basis is $0.5$. The detector dark count rates are below \SI{191}{\hertz} per detector at efficiencies higher than $80\%$ and a jitter lower than \SI{40}{\pico\second} \cite{Caloz2018}.

\begin{figure}
\includegraphics[ trim={0 0 0 0},clip, width = 0.5\columnwidth]{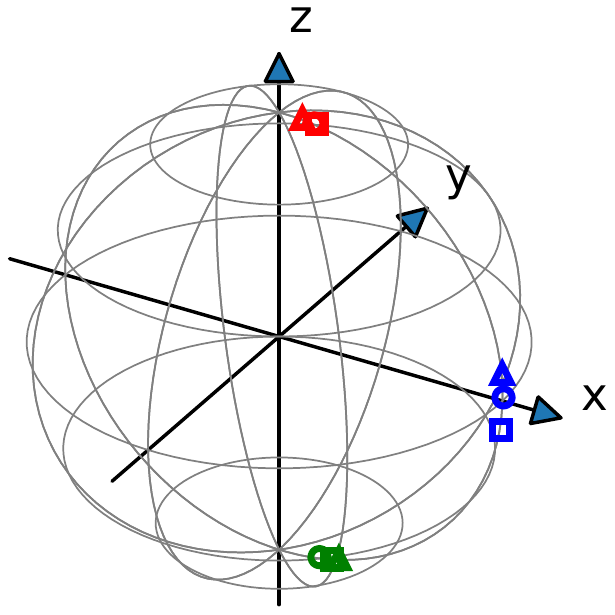}
\caption{\label{fig:results_polar} States prepared by Alice in the Poincaré sphere. The state $\ket{\psi_{0|k}}$ is marked in red, $\ket{\psi_{1|k}}$ in green and $\ket{\psi_{+|k}}$ in blue. The index $k$ is the value of the previous state. The triangles mark  states where $k  = 0$, the squares mark states where $k  = 1$ and the circles mark states where $k=+$.}
\end{figure}

Sources of errors are detector noise, misalignment of Alice's and Bob's bases and imperfect preparation of the states at Alice. Indeed, Alice generates the states
\begin{equation}\label{eq:states_with_spf}
\begin{aligned} 
\ket{\psi_{j}} = \cos\frac{\theta_j}{2}\ket{0} +  \sin\frac{\theta_j}{2}\ket{1}
\end{aligned}
\end{equation}
where $j = 0, 1$ or $+$. In practice, the angles $\theta_j$ deviate slightly from the ideal values $0,\pi$ and $\pi/2$, a limitation called state preparation flaws (SPF). In \autoref{fig:results_polar} we observe that the $\ket{\psi_0}$ and $\ket{\psi_1}$ are not perfectly orthogonal.  

We use a low-density parity-check (LDPC) code for error correction \cite{Elkouss2010}, implemented in the FPGA and offering a high throughput. For a quantum bit error rate (QBER) $Q_\text{Z}$ up to $3 \%$, the code rate is $2/3$, i.e. the leakage $\lambda_\text{EC}$ is one third of the sifted key. The used implementation corrects blocks of  $1944$~bits and one privacy amplification block consists of $4192$ error correction blocks.

The phase error rate $\phi_{Z}$ is estimated following the loss-tolerant quantum cryptography analysis \cite{Tamaki2014,Pereira2019}, taking into account the SPF. In the three state protocol, not only the events where Alice and Bob choose the X basis are used to estimate the phase error rate, but also the events where Alice and Bob choose the Z basis and the X basis, respectively. The measured SPF (see \autoref{tab:results_angles}) increase the number of detections of the state $\ket{+}$ when Alice sends a state in the Z basis, leading to an overestimated phase error rate. We can compensate for this increased number of events and obtain a better estimate of the phase error rate following the loss-tolerant security analysis \cite{Tamaki2014,Pereira2019}.

The number of single photon events in the Z basis, noted $s_{1,Z}$, and the single photon events needed to bound $\phi_{Z}$ is estimated using the decoy method \cite{Rusca2018}. After privacy amplification, Alice and Bob hold a secret key of length
\begin{align}\label{eq:sk_length}
l \geq& s_{1,Z}(1-h(\phi_Z)) - \lambda_\text{EC} \nonumber\\
& - 6\log_2(19/\epsilon_\text{sec}) - \log_2(2/\epsilon_\text{corr}).
\end{align}
The parameters $\epsilon_\text{sec} = \num{1e-9}$ and $\epsilon_\text{corr} = \num{1e-15}$ are the security and correctness parameters, respectively.

\begin{table*}[ht]
\centering
\begin{tabular}{c c c c c c}
\hline
Fiber length (km) & Attenuation (dB) & Sifted key rate (kbps)& $\phi_\text{Z}$ (\%) & $Q_\text{Z}$ (\%) & SKR (kbps)  \\ 
\hline
101.0 & 20.2 & 2320.2  & 3.67 & 1.93 & 392.7 \\
151.5 & 30.3 & 330.0 & 3.50  & 1.88 & 54.5 \\
\hline
\end{tabular}
\caption{\label{tab:results_skr} Measured secret key rate (SKR) and corresponding experimental parameters. $\phi_\text{Z}$ is the phase error rate and $Q_\text{Z}$ is the QBER Z. The given rates are averages over ten privacy amplification blocks.}
\end{table*} 


\autoref{tab:results_skr} shows the SKR and the corresponding QBER and phase error rate in the Z basis versus the quantum channel fiber length. At the distance of \SI{101.0}{\kilo\meter} our detectors are at the edge of saturation. These secret key rates do not take into account the correlations between succeeding pulses which we discuss in the following chapters. We achieved an average SKR of 54.5~kbps over ten privacy amplification blocks after \SI{151.5}{\kilo\meter} of single mode fiber. Compared to already existing implementations, this is an improvement of more than a factor of five at the same distance \cite{Frohlich2017,Wang2012, Takesue2007b}.
 
In addition to SPF discussed above, we can also have correlations between the states. In case of nearest neighbour correlations, the states take the form
\begin{equation}\label{eq:states_with_corr}
\begin{aligned} 
\ket{\psi_{j|k}} = \cos\frac{\theta_{j|k}}{2}\ket{0} +  \sin\frac{\theta_{j|k}}{2}\ket{1},
\end{aligned}
\end{equation}
where $k = 0,1$ or $+$ is the bit value of the previous state. The angle is now $\theta_{j|k} = \theta_j + \delta_{j|k}$. We say that correlations are present when the angles $\delta_{j|k}$ are not zero.  

We determine the correlations in the polarization of the pulses as follows:  Alice emits repeatedly a sequence of 32 states with the probabilities $p_\text{Z}^\text{A}/2$ for $\ket{\psi_0}$ and $\ket{\psi_1}$ and probability $p_\text{X}^\text{A}$ for $\ket{\psi_+}$. 
At Alice's quantum channel output (see \autoref{fig:setup}) we put a simple polarimeter consisting of a rotatable quarter-wave plate (QWP) in a U-bench, a linear polarizer and a SNSPD. This configuration allows us to infer the Stokes parameters of each pulse in the sequence \cite{Schaefer2007}. The average of a state over all previous states is given by
\begin{equation}
\begin{aligned} 
\ket{\psi_{j}} = 
\frac{
	\sum_{k}\ket{\psi_{j|k}}
}{
	|\sum_{k, l}\braket{\psi_{j|k}|\psi_{j|l}}|^2
}
\end{aligned}
\end{equation}
This average only depends on $\theta_{j}$. The correlation angles $\delta_{j|k}$ are retrieved from the overlap of the measured states $\ket{\psi_{j|k}}$ with $\ket{\psi_{j}}$.

\begin{table}[ht]
\centering
\begin{tabular}{l c c c}
\hline
& $\ket{\psi_{0}}$ & $\ket{\psi_{1}}$ & $\ket{\psi_{+}}$ \\ 
\hline
Average angle $\theta_{j}$ (\si{\degree}) & 8.0 & 165.6 & 90.0 \\
Max. deviation $\max_{k}\delta_{j|k}$(\si{\degree}) & 6.3 & 6.9 & 8.0 \\
\hline
\end{tabular}
\caption{\label{tab:results_angles} Measured angles of the average states and the maximum deviation from this angle. }
\end{table}
The measured states are shown on the Poincaré sphere in \autoref{fig:results_polar} and the values of $\theta_{j}$ and of the maximum deviations angles $\max_{k}\delta_{j|k}$ can be found in \autoref{tab:results_angles}. The reference frame for the measurement was chosen such that the average angle $\theta_+$ is exactly $\SI{90}{\degree}$. In \autoref{fig:results_polar} it can be seen that the states show correlation with the pulse which was sent before. This is mainly due to the the distortion introduced by the limited bandwidth of the electronic supply chain of the PM.

There exists a security proof considering correlation between the states \cite{Pereira2019b}, however it requires the angles $\delta_{j|k}$ to be of the order of  $\SI{0.1}{\degree}$ which is challenging to achieve in practice. Fluctuations by this angle would correspond to a QBER of $\num{3e-6}$. More theoretical work is needed to obtain a proof with a tighter bound on the secret key rate. 


\begin{figure}
\includegraphics[ trim={0 0 0 0},clip, width = 0.9\columnwidth]{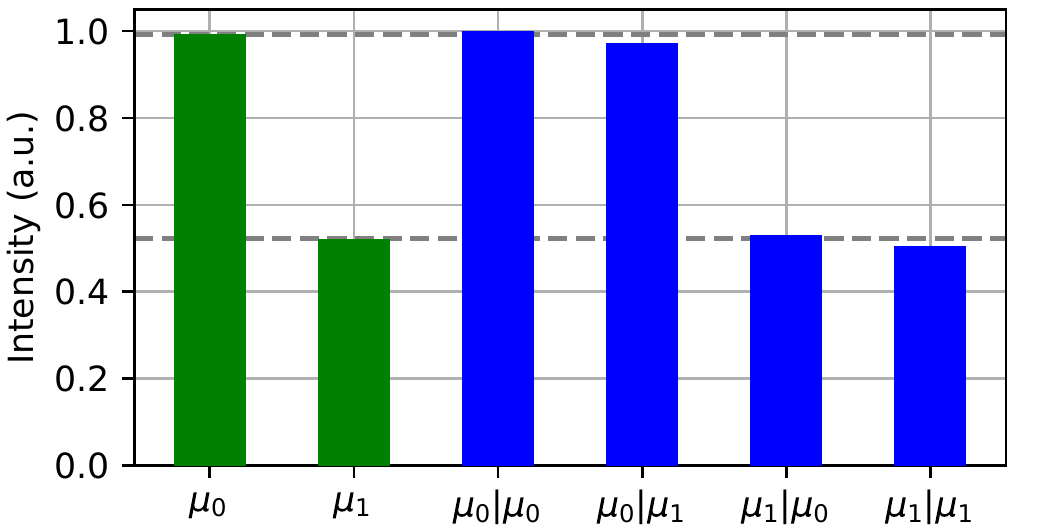}
\caption{\label{fig:decoy_characterization} Intensity of the decoy states. The first two bars show the mean intensity of the signal ($\mu_0$) and decoy ($\mu_1$) state. The four last bars show the mean intensity of a state conditioned on the state sent before.}
\end{figure}

To characterize the correlations in the decoy intensity we use Alice's setup (grey box in \autoref{fig:setup}) and prepare repeatedly a sequence of 32 pulses containing all the four nearest-neighbour combinations. The sequence is measured with an optical oscilloscope at the output of Alice. We measured the relative intensities of the decoy and signal state conditioned on which state was sent before. The results are depicted in \autoref{fig:decoy_characterization}. We see a slight relative change in intensity of up to $3\%$, depending on what was sent before.

There exists a security analysis which takes into account only independent and identically distributed fluctuations of the decoy intensities \cite{Mizutani2015}. However, following this security proof, the finite key correction terms go with the square root of the number of pulses sent by Alice. In contrast to the case of uncorrelated sources where the finite key correction terms scale with the square root of the number of pulses detected by Bob. This means that huge block sizes of the order of $10^{10}$ are required if we don't want the maximum distance to be limited by this effect. There is a technique based on post-processing operations which can be applied to correlated decoy intensities \cite{Yoshino2018}, but further work is needed to evaluate the impact of these correlations.


In order to determine to which degree the phase between two succeeding laser pulses is correlated, we employ a simple model of a gain-switched laser source. We assume that a pulse has a perfectly correlated phase with respect to the previous pulse with probability $p_\text{c}$ or else the phase is completely arbitrary. Under this assumption, we measure the degree of phase correlation $p_\text{c}$ using a fiber Michelson interferometer with a tunable arm with length $\Delta d$ which is needed to maximize the overlap of two subsequent pulses. To estimate the interferometer's imperfections, we measure the fringe visibility $V_\text{CW}(\Delta d)$ when operating the laser source in continuous wave mode. Then we measure the fringe visibility $V_\text{pulsed}(\Delta d)$ operating the laser source with the same parameters as for the QKD setup. The degree of phase correlation is then estimated by
\begin{align}\label{eq:phase_correlation}
p_\text{c}^* &= \max_{\Delta d}\frac{V_\text{pulsed}(\Delta d)}{V_\text{CW}(\Delta d)}
\end{align}
optimizing $\Delta d$.  By doing so, we would overestimate $p_\text{c}$ if the  coherence length of the laser source in continuous wave mode wasn't sufficient. We find a very small value of $p_\text{c}^*  = 0.0019$. 

What is the influence of an imperfect phase randomization on the security of the protocol? According to our model, this means that a proportion $p_\text{c}^*$ of pulses are potentially insecure and Eve could use the phase coherence to obtain a bit more information. In the worst case scenario each insecure pulse contributes a bit to the final key and Eve can infer this bit due to non-random phase between this pulse and the adjacent pulses. Therefore we can avoid this loophole by reducing the key length by a factor $1-p_\text{c}^*$ before privacy amplification. Given the measured value, we can conclude that the degree of phase correlation has a negligible effect on the SKR. Still, we take it into account by multiplying \autoref{eq:sk_length} by $1-p_\text{c}^*$.


We demonstrated the implementation of a BB84 protocol at \SI{5}{GHz} pulse repetition rate and established high SKRs over \SI{101.0}{\kilo\meter} and \SI{151.5}{\kilo\meter} of fiber. We used state-of-the-art secret key distillation   taking account finite key analysis and SPF. 
However, we identified three potential security loopholes opening in particular at high clock rates. These are correlations in the polarization state preparation, correlations between the mean photon numbers of succeeding pulses and imperfect phase randomization of the laser source. While we verified experimentally that these effects are reasonably small, we realize that with the current security proofs the impact on the secret key rates and distance can be important, in particular for the correlations in the state preparations. Arguably, this is an issue for all QKD systems also at lower repetition rate. Additional theoretical analysis is clearly needed here in future. 


We thank Margarida Pereira,  Kiyoshi Tamaki and Marcos Curty for the valuable discussions and Rapha\"el Houlmann and C\'edric Vulliez for programming the FPGAs. This work was financially supported by the Swiss NCCR QSIT. Davide Rusca thanks the EUs H2020 program under the Marie Sk{\l}odowska-Curie project QCALL (No. GA 675662) for financial support.


The data that support the findings of this study are available from the corresponding author upon reasonable request.

\bibliography{bibliography}
\end{document}